\documentclass[tradiabstract,referee]{aa} 
\usepackage{graphicx}
\usepackage{natbib}
\def\simgreat{\mathbin{\lower 3pt\hbox
     {$\rlap{\raise 5pt\hbox{$\char'076$}}\mathchar"7218$}}}
\def\simless{\mathbin{\lower 3pt\hbox
     {$\rlap{\raise 5pt\hbox{$\char'074$}}\mathchar"7218$}}}

\newcommand{\Lsun} {L$_{\odot}$}
\newcommand{\Msun} {M$_{\odot}$}
\newcommand{\Rsun} {R$_{\odot}$}

\begin{document}

\title{PAH destruction and survival in the disks of T Tauri stars}
\titlerunning{PAH destruction and survival in T Tauri disks}

\author {R.~Siebenmorgen\inst{1} \and E.~Kr\"ugel\inst{2}}
\institute{
        European Southern Observatory, Karl-Schwarzschildstr. 2,
        D-85748 Garching b. M\"unchen, Germany
\and
        Max-Planck-Institut f\"ur Radioastronomie, Auf dem H\"ugel 69,
        Postfach 2024, D-53010 Bonn, Germany 
}
\offprints{rsiebenm@eso.org}
\date{Received Month XX, 2009 / Accepted Month XX, 20XX}

\abstract {In Spitzer observations of Tauri stars and their disks, PAH
  features are detected in less than 10\% of the objects, although the
  stellar photosphere is sufficiently hot to excite PAHs.  To explain
  the deficiency, we discuss PAH destruction by photons assuming that
  the star has beside its photospheric emission also a FUV, an EUV and
  an X--ray component with fractional luminosity of 1\%, 0.1\% and
  0.025\%, respectively. {As PAH destruction process we consider
    unimolecular dissociation and present a simplified scheme to
    estimate the location from the star where the molecules become
    photo-stable. We find that soft photons with energies below $\sim
    20$\,eV dissociate PAHs only up to short distances from the star
    ($r < 1$\,AU); whereas dissociation by hard photons (EUV and
    X--ray) is so efficient that it would destroy all PAHs (from
    regions in the disk where they could be excited). } As a possible
  path for PAH survival we suggest turbulent motions in the disk.
  They can replenish PAHs or remove them from the reach of hard
  photons.  For standard disk models, where the surface density
  changes like $r^{-1}$ and the mid plane temperature like $r^{-0.5}$,
  the critical vertical velocity for PAH survival is proportional to
  $r^{-3/4}$ and equals $\sim$5\,m/s at 10\,AU which is in the range
  of expected velocities in the surface layer.  The uncertainty in the
  parameters is large enough to explain both detection and
  non-detection of PAHs.  Our approximate treatment also takes into
  account the presence of gas which, at the top of the disk, is
  ionized and at lower levels neutral. }


\maketitle

\section{Introduction}

Infrared emission bands of PAHs can be used as a probe of the UV
environment.  They are commonly seen in the ISM, but also in young
stellar objects such as Herbig Ae/Be stars (Waelkens et al. 1996,
Siebenmorgen et al. 2000, Meeus et al. 2001, Peeters et al. 2002, van
Boekel et al. 2004).  The observed emission can be explained in models
of an irradiated disk (Habart et al. 2004, Visser et al., 2007,
Dullemond et al. 2007a).

ISO also looked at a few of the much fainter T Tauri stars but without
a clear PAH detection (Siebenmorgen et al. 2000).  In the Evans et
al.~(2003) legacy program which employs the more sensitive Spitzer
Space Telescope (SST), 3 out of 38 T Tauri stars show PAH features
(Geers et al.~2006).  This corresponds to a detection rate of only 8\%
in contrast to almost 60\% in Herbig Ae/Be stars (Acke \& van den
Ancker 2004).  Similarly low rates for T Tauri stars are found by
Furlan et al.~(2006) who present 111 SST spectra in the Taurus-Auriga
star forming region and speculate that the absence of PAH resonances
is due to the much weaker UV field compared to Herbig Ae/Be stars.
Geers et al.~(2009), on the other hand, argue that the PAHs are simply
under-abundant relative to the ISM.  They also find that variations of
the disk geometry, such as flaring or gaps, have only a small effect
on the strength of the PAH bands.  Clearing out gas and dust by planet
formation inside the disk could effectively remove PAHs. Indeed, inner
gaps in disks are observed at radii between 40--60\,AU and at
wavelengths between 20--1000$\mu$m where the emission is dominated by
large grains (Brown et al. 2008, Geers et al. 2007b).  However, in
cases where PAH emission is resolved, it is extended up to 15--60\,AU,
without sub-structure and inside the inner gap region (Geers et
al. 2007a).  The spatial extent of the PAH emission is also similar
for T Tauri and Herbig Ae/Be stars.  We therefore suggest that PAH
removal by radiative destruction is dominant.

Present radiative transfer models of the PAH emission from dusty disks
consider only the stellar radiation field and no additional EUV or
X--ray component (Habart et al.~2004, Geers et al.~2006, Visser et
al.~2007, Dullemond et al.~2007).  Their hard photons could, according
to laboratory experiments (Ruhl et al. 1989, Leach et al. 1989a,b,
{Jochims et al. 1994}) and theory (Omont 1986, {Tielens 2005},
Rapacioli et al. 2006, { Micelotta et al. 2009}), destroy PAHs.  We
discuss below their impact on the PAH abundance in the disks of T
Tauri stars.

\section{Radiation components of T Tauri stars} \label{rad.sec}

\begin{figure}
\begin{center}
\includegraphics[angle=0,width=12cm]{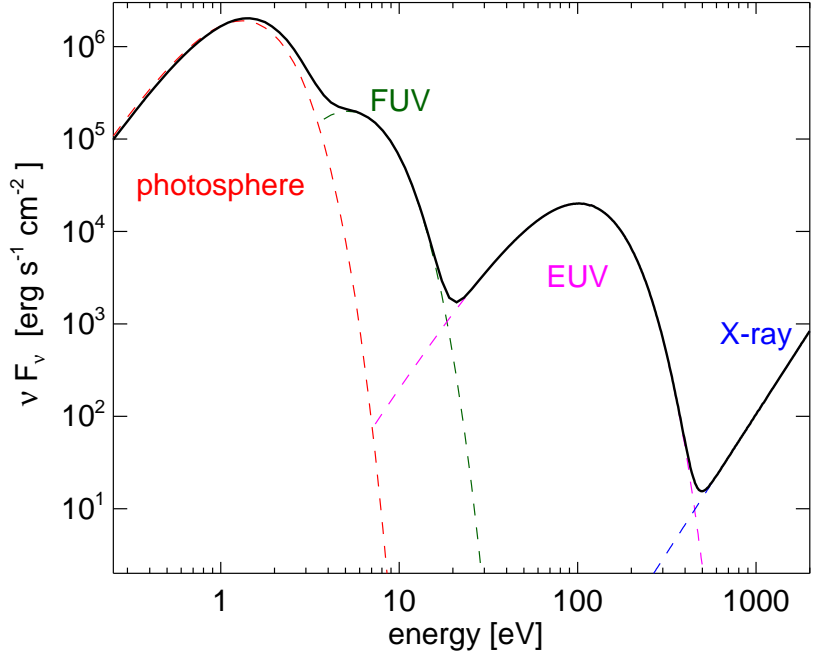}
\end{center}
\caption{The spectral energy distribution of our T Tauri model star at
1\,AU without foreground extinction (Eq.~\ref{flux.eq}).  The absolute
luminosities of the components are given in Table~\ref {spec.tab}.
\label{fnu.ps}}
\end{figure}

Our T Tauri model star has a total luminosity $L_*=2$\,\Lsun.  Its radiation
consists of a photospheric, a FUV, an EUV and an X--ray component.  Their
parameters are listed in Table~\ref{spec.tab} and are very similar to those
proposed by Gorti \& Hollenbach (2008).  The total spectrum is displayed in
Fig.\ref{fnu.ps}.  We point out that the FUV and EUV radiation are
observationally poorly constrained.

The photosphere supplies most of the luminosity whereas the FUV, EUV and X--ray
radiation are much weaker and believed to originate from accretion onto the star
and from chromospheric and coronal activity.  The photosphere, the FUV and EUV
component are approximated by blackbodies.  We assume 4000\,K for the
photosphere and, following Stahler et al.~(1980) and Calvet \& Gullbring (1998),
15000\,K for the FUV (pre-shock) and $\sim$3$\times 10^5$\,K for the EUV
emission (post-shock region).

The strength of the FUV and EUV radiation is determined by the accretion
luminosity which we approximate by $L_{\rm acc} = G M_* \dot{M} /R_*$.  If $R_*
= 2$\,\Rsun\ and $M_* = 1$\,\Msun\ are the radius and mass of the star, an
accretion rate $\dot{M}= 10^{-9}$ \Msun\ yr$^{-1}$ (Akeson et al. 2005) yields
$L_{\rm acc} \sim 0.01\, L_*$.  Higher values ($L_{\rm acc}/L_* \simgreat 0.1$),
but with a large spread, are derived by Muzerolle et al.~(1998, 2003) from
hydrogen emission lines.  However, as we show in section~\ref{surv}, such
stronger fluxes have little influence on the stability analysis of PAHs.

\begin{table*}[htb]
  \caption{\label{spec.tab} The four radiation components of our T Tauri model 
    star. }
\begin{center}
\begin{tabular}{|l | l | c | c | c | c |}
\hline
  &   &(1) &(2)  & (3) & (4)\\ 
\hline
& &   & &  & \\ 
$i$ & component &$L/L_*$ & spectrum & $h\overline{\nu}_{\rm em}$ & $F_{10}$ \\ 
& & & & (eV) & (erg s$^{-1}$cm$^{-2}$) \\
\hline
1& photosphere & 0.99   &4000\,K  \ BB              & 0.9 & 30\,000 \\
2& FUV        & 0.01   &15000\,K \ BB                 & 3.5 & 300 \\
3& EUV        & 0.001  &$3\times 10^5$\,K \ BB    & 70 & 30 \\
4& X--rays ($h\nu < 2$\,keV) & $2.5\times 10^{-4}$& $\propto\nu^2$& 1330 & 10 \\
\hline
\end{tabular}
\end{center}

\hspace{2cm} (1) fractional luminosity, 

\hspace{2cm} (2) spectral shape (BB = blackbody),

\hspace{2cm} (3) mean energy of emitted photons, $h\overline{\nu}_{\rm
{em}}$,

\hspace{2cm} (4) approximate flux at 10\,AU, $F_{10}$.
\end{table*}

Preibisch et al.~(2006) establish from Chandra observations (0.5\,--\,8\,keV) a
relation between the X--ray luminosity, $L_x$, and the total luminosity $L_*$
confirming the ROSAT results of Sterzik \& Schmitt (1997).  The ratio $L_x/L_*$
is similar in rapidly rotating main-sequence stars and non-accreting T Tauri
stars ($\sim$10$^{-3}$), but systematically lower by a factor $\sim$4 in
accreting T Tauri stars (Preibisch et al. 2006).  Interestingly, in Herbig Ae/Be
stars $L_x/L_*$ is much smaller ($\sim$$10^{-7} \ldots \ 10^{-5}$, Stelzer et
al.~2006) with values comparable to the Sun.  The solar X--ray luminosity in the
0.1--2.4\,keV ROSAT passband lies during a solar cycle in the range $10^{-6.8}
\simless L_x/L_* \simless 10^{-5.7}$ and is typical for G stars (Judge et
al. 2003).
 
X--ray fluxes are generally variable on timescales of hours to weeks and weaken
during the evolution of the T Tauri star.  For example, half of the sources in
the Taurus molecular cloud detected by XMM/Newton (0.3 -- 7.8\,keV) show
variations, more at hard ($>0.5\,$keV) than at soft energies, and a quarter of
them display flares (Stelzer et al. 2007), about once a week and lasting for a
few hours.  In a strong flare, more than $10^{35}$\,erg are emitted and $L_x$
can reach 1\% of the total luminosity.  We assume up to 2\,keV a power law
spectrum $\propto \nu^2$ (G\"udel et al.~2007) and neglect harder radiation
because the emission then steeply declines ($\propto \nu^{ - 3}$).

Let $L_i$ be the frequency-integrated luminosity of the radiation component $i$
(see Table~\ref{spec.tab}) and $L_{i,\nu}$ its spectral luminosity such that
$L_i = \int L_{i,\nu}\, d\nu$.  Dropping for convenience the index $i$, the flux
(of component $i$\,) at a distance $r$ is
\begin{equation}\label{flux.eq}
F_\nu = {L_\nu e^{-\tau_\nu} \over 4\pi r^{2}}
\end{equation}
where we included a screening factor $e^{-\tau_\nu}$ to account for
foreground absorption (by dust and gas).  If $\kappa_\nu$ denotes the
absorption cross section per carbon atom, a PAH of $N_c$ carbon atoms
absorbs in one second (from component $i$)
\begin{equation}\label{N_gamma}
N_\gamma = N_c \int {F_\nu \kappa_\nu\over h\nu} \,d\nu 
\end{equation}
photons of total energy
\begin{equation}\label{E_abs}
E_{\rm abs} =  N_c \int F_\nu \kappa_\nu \, d\nu
\end{equation}
The inverse of $N_\gamma$ is the average time between two absorption events,

\begin{equation} \label{t_abs} 
t_{\rm abs} = N_\gamma^{-1} 
\end{equation}
The mean photon energy equals
\begin{equation}\label{hnu_mean}
 h\overline{\nu} = {\int F_\nu \kappa_\nu d\nu 
     \over \int {F_\nu  \kappa_\nu\over h\nu}  d\nu }
\end{equation}

\section{Cross sections}

As the light from the star enters the disk, it is attenuated by gas and dust.
The absorption cross section of gas depends on the ionization stage of the atoms
which is determined by the balance between recombination and photo-ionization.
By far the most important atoms are, of course, hydrogen and helium with
ionization potentials of 13.6\,eV and 24.6\,eV, respectively.  Because the
recombination rate is proportional to the square of the gas density which is
high in the disk (section 5), the gas is ionized only in a thin surface layer
($A_{\rm V} <$ 0.001\,mag, section 2.6 of Gorti \& Hollenbach 2008).  We use
atomic cross sections of Morrison \& McCommon (1982) and Balucinska--Church \&
McCommon (1992) and solar element abundances.

The dust cross sections are taken from the model of Kr\"ugel ({
  2006}) which describes standard dust.  For X--rays, the absorption
efficiency calculated from Mie theory must be corrected downwards.
Hard photons can eject electrons from the grain and as these carry
away kinetic energy, only part of the photon energy is deposited in
the dust particle.  The threshold, $E_{\rm t}$, above which such a
correction is necessary depends on the grain size; details are given
in Dwek \& Smith (1996).  For a 10\,\AA\ graphite particle, $E_{\rm t}
\sim 100$\,eV and the reduction factor is roughly proportional to
$\nu^{-1}$.

\begin{figure*}
\begin{center}
\includegraphics[angle=0,width=12cm]{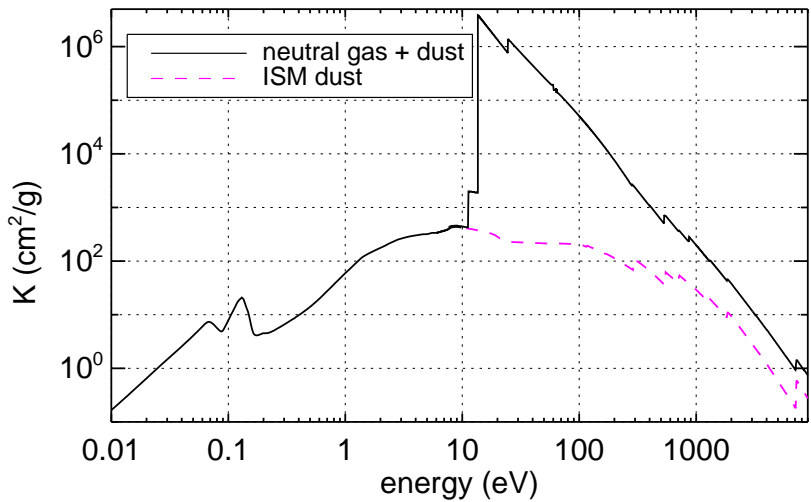}
\end{center}
\caption{The mass extinction coefficient per gram disk material when the gas is
  neutral (Morrison \& McCommon 1983); the gas--to--dust mass ratio equals
  130.\label{sigt.ps}}
\end{figure*}

The absorption coefficient of dust, $K_{\rm d,\lambda}$, and of neutral gas plus
dust, $K_\lambda = K_{\rm gas,\lambda} + K_{\rm d,\lambda}$, both per gram of
disk material, are plotted in Fig.~\ref{sigt.ps} for a dust-to-gas mass ratio of
1:130.  Note that at the ionization threshold of hydrogen, $K_{\rm gas}$ is
almost 10$^4$ times greater than $K_{\rm d}$.

With respect to the absorption cross section of PAHs, we assume $\kappa_\nu =
7\times 10^{-18} $\,cm$^2$ per carbon atom when $h\nu < 13.6$\,eV and scale
$\kappa_\nu$ at higher energies to follow the values of a graphite sphere of
10\,\AA\ radius (Dwek \& Smith 1996).  The maximum wavelength (in \AA) for PAH
excitation is $\lambda_{\rm max,PAH} =1630 + 370\sqrt{N_c}$ (Schutte et
al.~1993) resulting in a minimum photon energy of 2.3\,eV for a PAH with
$N_c=100$ carbon atoms.

\subsection{PAH emission}

As the PAHs are transiently heated, their excitation is usually
treated statistically.  Following Guhathakurta \& Draine (1989), let
$P(T)\,dT$ be the probability of finding in a large ensemble of PAHs
in a steady state an arbitrary PAH in the temperature interval $[T,
T+dT]$.  The temperature distribution function $P(T)$ is calculated in
this method from a transition matrix $(A_{fi})$.  If $K_\nu$ denotes
the PAH absorption cross section, the matrix element $A_{fi}$
referring to dust heating from an initial enthalpy bin centered at
$U_i$ to a final one centered at $U_f$ and of width $\Delta U_f$ is,
for a mono--chromatic flux, equal to

\begin{equation}
\nonumber
A_{fi} = 
\left\{ \begin{array}{ccl} 
\displaystyle{K_\nu F \over h\nu}  & : &\ \mbox{if} \ |U_f-U_i-h\nu| \le
{1\over 2} \Delta U_f  \\
0   & : & \ \mbox{else} \nonumber   \end{array}  
\right.
\end{equation} 

\begin{figure*}
\begin{center}
\includegraphics[angle=0,width=12cm]{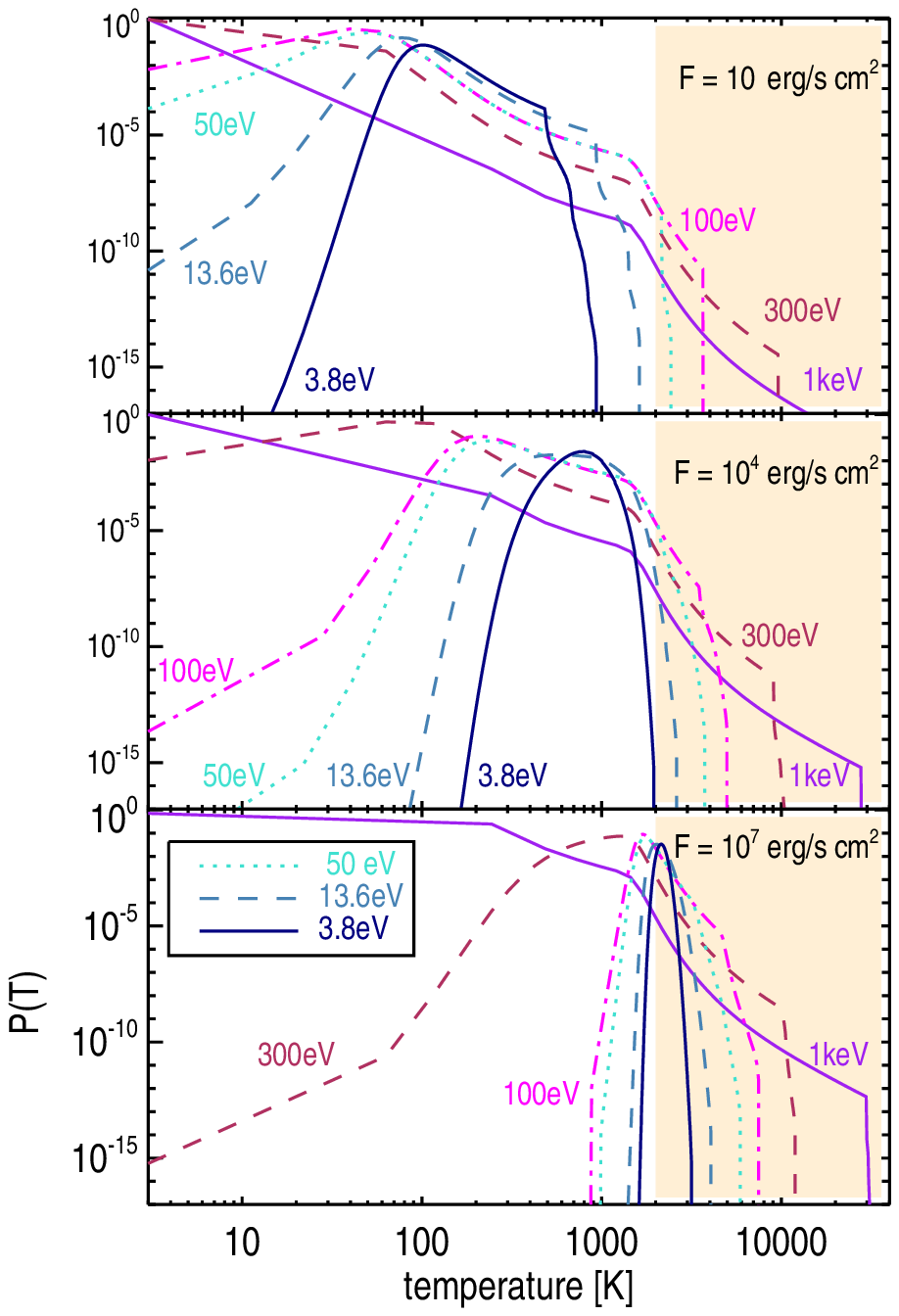}
%\vspace*{-1.0cm}
\end{center}
\caption{\label {Fig3.ps} The temperature distribution $P(T)$ of a PAH
  with 100\,C atoms exposed to mono--chromatic radiation with $h\nu=
  3.8, 13.6, 50, 100, 300$\,eV and 1\,keV.  This set includes the mean
  photon energies of the four radiation components of the T Tauri
  star.  The fluxes range from (top to bottom) $F =10$ to $10^7$ erg
  s$^{-1}$ cm$^{-2}$. Shaded area marks temperatures above the
  sublimation temperature of graphite. }
\end{figure*}

\begin{figure}
\begin{center}
\includegraphics[angle=0,width=12cm]{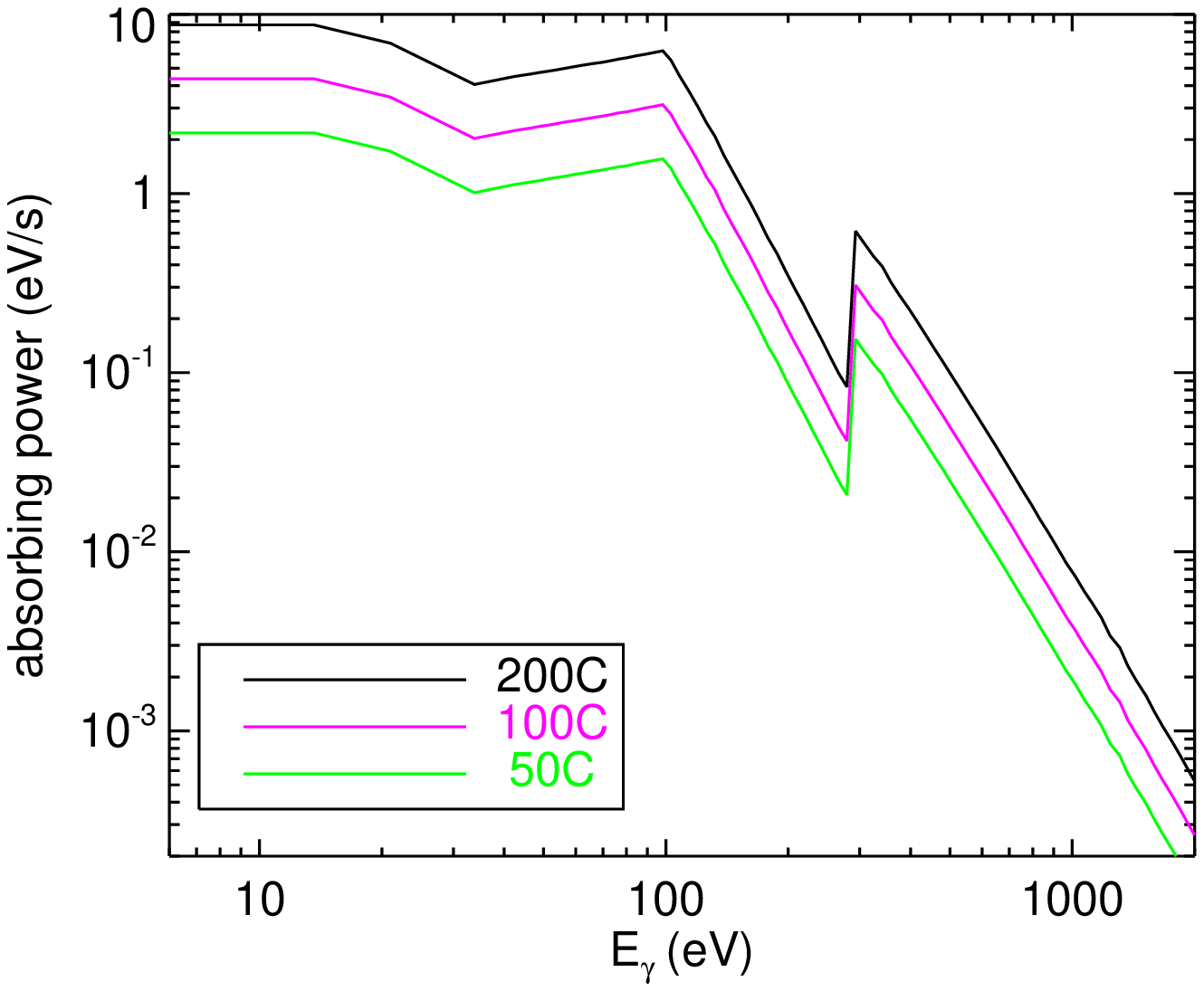}
\end{center}
\caption{The power $W = N_{\rm c} \ \kappa \ F$ absorbed by one PAH, with
number of C atoms $N_{\rm C}$ as indicated, in a mono--chromatic flux
$F =10^4$\,erg s$^{-1}$\,cm$^{-2}$ as a function of photon energy
$E_{\gamma}=h\nu$.
\label {2Proba3.ps}}
\end{figure}

{Examples of $P(T)$ are displayed in Fig.~\ref{Fig3.ps} for
  mono--chromatic fluxes which cover almost the entire range
  encountered anywhere in the disk in terms of intensity, the flux
  ranges from $F=10$ to $10^7$\,erg\,s$^{-1}$cm$^{-2}$, as well as
  hardness, the photon energy is between $h\nu = 3.8$\,eV and 1keV}.
The approximate unattenuated fluxes of the four radiation components
at a distance of 10\,AU are listed in Table~\ref{spec.tab}.  Note that
when $F = n_\gamma h\nu$ is constant, the number of photons $n_\gamma$
decreases as the photon energy $h\nu$ goes up.  {The power absorbed by
one PAH is almost independent of the photon energy as long as
$h\nu\simless 100$\,eV (see Fig.~\ref {2Proba3.ps}).}  By and large,
when $h\nu$ is fixed and $F$ increases, the curves in
Fig.~\ref{Fig3.ps} narrow and move to the right {towards higher
  temperatures. When the radiation field is weak
  ($F=10$\,erg\,s$^{-1}$cm$^{-2}$) and the photons are soft
  ($h\nu=3.8$\,eV), the PAH absorbs about one photon per day and there
  is plenty of time to cool down. In this case, the PAH virtually
  never exceeds the sublimation temperature $T_{\rm s}$, (at which
  solid carbon gasify). For a wide pressure range ($10^{-1} -
  10^{-7}$\,dyn \,cm$^{-2}$), $T_{\rm s} \sim 2000$\,K for graphite
  (CRC Handbook of Chemistry \& Physics 2005, Salpeter et al. 1977).
  In Fig.~\ref{Fig3.ps} we highlight the area where the temperature is
  above $T_{\rm s}$.  For a PAH exposed to a strong radiation field
  ($F=10^7$\,erg\,s$^{-1}$cm$^{-2}$), about a dozen of soft photons
  are absorbed within a cooling time and the temperature distribution
  function becomes very narrow around 2000\,K and sublimation is
  likely.  For hard photons ($h \nu \geq 50$\,eV), the PAH undergoes,
  independent of the strength of the radiation field, extreme
  temperature excursions. This indicates that PAH become
  photo--unstable either by absorption of a single hard photon, or by
  soft photons if there are many of them.}

\section{PAH destruction}

The abundance of PAHs is determined by the competition between
formation and destruction processes under the specific environmental
conditions.  Underlying processes are discussed, for example, by Omont
(1986), Voit (1992), or recently by Micelotta et al.(2009a).  {
  Here we only consider PAH destruction by photons and generally
  assume that PAH formation is negligible.  After photon absorption, a
  highly vibrationally excited PAH may relax through emission of IR
  photons or, if sufficiently excited, lose atoms.  The latter process
  is called unimolecular dissociation and is discussed for
  interstellar PAHs by Allamandola (1989), Leger et al. (1989), Le
  Page et al. (2003), Rapacioli et al.~(2006), and Micelotta et
  al.~(2009b).  Laboratory studies of PAH dissociation which can be  
  applied to astrophysical situations are rare (Jochims et al. 1994).
  The photo--chemistry of PAHs is reviewed by Tielens (2005, 2008). }

\subsection{Procedure}

In the disks of T Tau stars, the PAH abundance depends obviously on
place and on time as the disk evolves.  There is no general solution
to the problem and to extract numbers, we have to radically simplify
it. {We wish to find some estimate of the location where PAHs
  become stable against photo--destruction. } To derive a procedure,
we recall that although after absorption of an energetic photon its
energy is immediately distributed over all available vibrational modes
(Allamandola et al., 1989), the excitation of a particular atom
fluctuates and occasionally it is pushed into the continuum and leaves
the PAH.  Quantitatively, {the unimolecular dissociation can be
  written in Arrhenius form. In a classical description, an atom of
  critical (Arrhenius) energy $E_{\rm 0}$ detaches from a PAH of peak
  temperature $T_{\rm p}$ if the dissociation time
\begin{equation}\label{t_dis}
t_{\rm dis} \sim \nu_0^{-1} e^{E_{\rm 0}/kT_{\rm p}}
\end{equation}

is shorter than the cooling time $t_{\rm cool}$.  A characteristic
value for the vibrational frequency is $\nu_0 \sim 10^{13}$\,s$^{-1}$.
The ``atom'', which may besides H or C also be an atomic group like
C$_2$H$_2$, needs the time $t_{\rm dis}$ to overcome the critical
internal barrier, $E_{\rm 0}$, which is similar but not identical to
the chemical binding energy. Micelotta et al. (2009b) quote $E_0$ of
3.2eV for H-loss, 4.2eV for C$_2$H$_2$, 7.5eV for pure C loss and
9.5eV for C$_2$.  For the ISM they find $E_0 = 4.6$\,eV and a somewhat
larger value for PDR. The inverse of the dissociation time } is the
probability that a certain atom leaves the PAH per unit time.

The exponential term $e^{E_{\rm 0}/kT_{\rm p}}$ in Eq.(\ref{t_dis})
increases very rapidly as $T$ falls and meaningful values (i.e.~not
too large ones) of $t_{\rm dis}$ are obtained only if $T_{\rm p} >
1500$\,K.  Atoms will only detach when $t_{\rm dis} < t_{\rm cool}$.
As the cooling time at these temperatures is for astrophysical
applications of order 1\,s, independent of the PAH size, the
dissociation criterion reads
\begin{equation}\label{t_dis2}
t_{\rm dis} \ \simless \ 1 \, {\rm s}
\end{equation}

It leads to a minimum temperature for destruction
\begin{equation}
T_{\rm dis} =  {E_{\rm 0} \over k \ln \nu_0} 
\end{equation}

Assuming $E_{\rm 0} \sim 5$\,eV, one gets $T_{\rm dis} \simeq
2000$\,K.  {At this high temperature, the internal energy of a PAH
  is reasonably well approximated by $3N_ckT_{\rm dis}$ when also
  taking the presence of H--modes into account.} The minimum
temperature $T_{\rm dis}$ is related to a minimum energy input $\Delta
E$.  The PAH is therefore unstable to photons with

\begin{equation}\label{Stabil2}
\Delta E \ge h\nu_{\rm c} = 3N_ckT_{\rm dis} = {3 \over \ln\nu_0} N_c
E_{\rm 0} \simeq 0.1 N_c E_{\rm 0}
\end{equation}
or when the number of carbon atoms

\begin{equation}\label{Nc_dis}
  N_c \ \le \ {2  \Delta E\over {\rm [eV]}}
\end{equation}

{Micelotta et al. (2009b) find that a PAH with $N_c=50$ requires
an internal energy of about $\Delta E = 24$\,eV to dissociate, which
agrees with the above estimate (Eq.~\ref{Nc_dis}).

The minimum energy input required for dissociation can either be
delivered by absorption of {\it i)} many soft photons, with a total
energy $E_{\rm {abs}} \geq \Delta E$ (Eq.~\ref{E_abs}), or {\it ii)}
by a single hard photon, with energy $h\nu \geq \Delta E$}. If a
photon heats the PAH to a peak temperature much above $T_{\rm dis}$,
more than one atom will detach.  The first expulsion occurs
momentarily ($t_{\rm dis} \ll 1$\,s).  It consumes the energy $E_{\rm
  0}$ plus some kinetic energy $E_{\rm kin}$ for the liberated atom.
The new PAH temperature follows from

\begin{equation}
h\nu - E_{\rm 0}-E_{\rm kin} = 3(N_c-1)kT 
\end{equation}
This happens $x$ times until $T$ has dropped to $T_{\rm dis}$,

\begin{equation}
h\nu - x(E_{\rm 0} + E_{\rm kin}) = 3(N_c-x)kT_{\rm dis} 
\end{equation}
With $E_{\rm kin} \sim 0.5$\,eV, we estimate that the total number of freed
atoms is

\begin{equation}\label{x_Atome}
x = {h\nu - 3N_ckT_{\rm dis}\over E_{\rm 0} + E_{\rm kin} - 3kT_{\rm dis} }
\simeq {h\nu \over 5\, {\rm [eV]}} - {N_c \over 10} \;
\end{equation}
For $N_c=100$, an average EUV photon ejects nine atoms and an X-ray photon
destroys the whole PAH (column (5) in Table~\ref{dis.tab}).

\begin{figure*}
\begin{center}
\includegraphics[angle=-90,width=12cm]{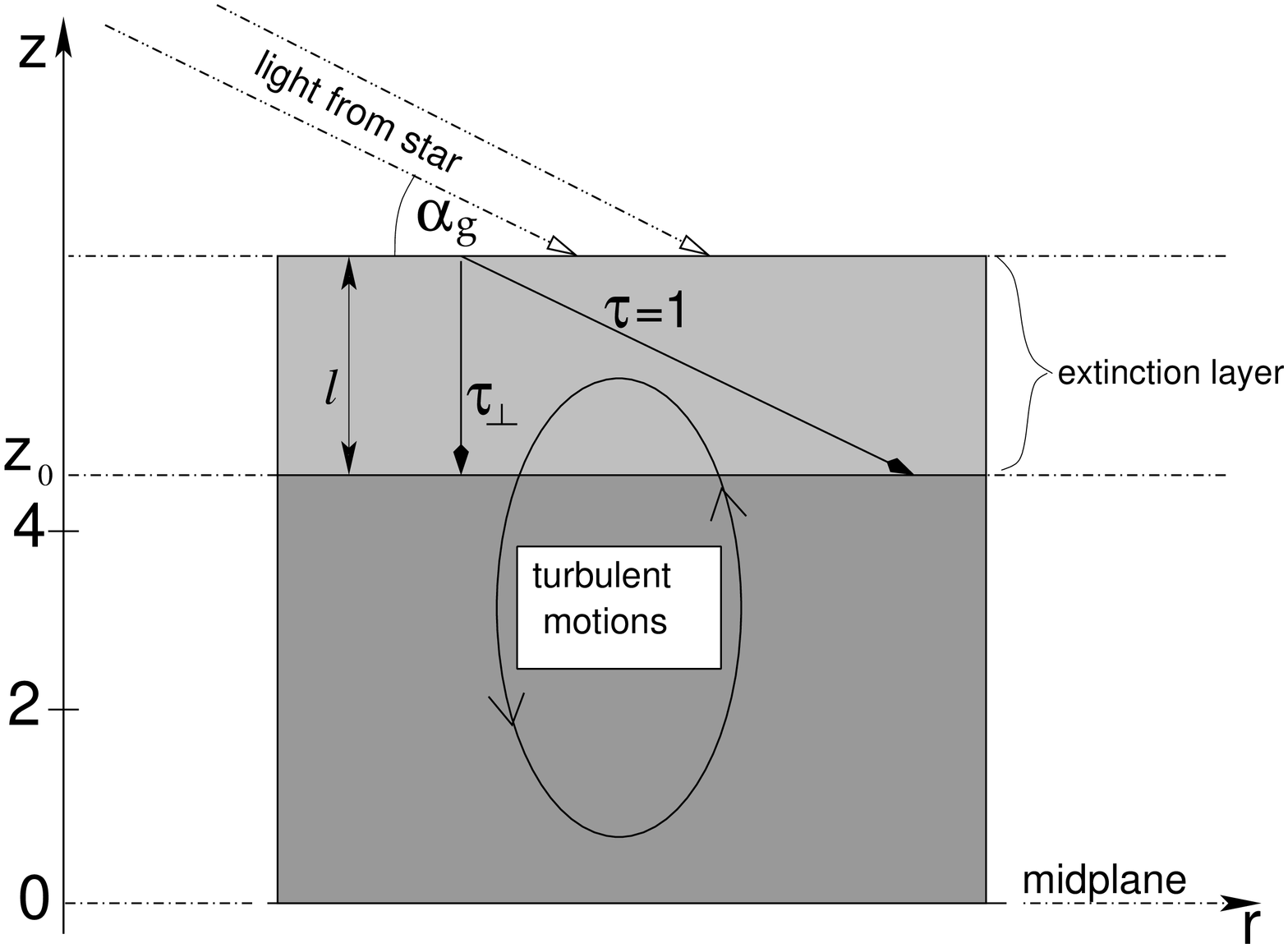}
\end{center}
\caption{Of each radiation component, $\sim$90\% is absorbed in what
  we call the extinction layer.  The optical depth from its bottom to
  the star is one and in vertical direction equal to the grazing angle
  $\alpha_{\rm g}$.  The height of its lower boundary, $z_0$, declines
  with radius, but its geometrical thickness is rather constant ($\ell
  \sim 0.5$\,H, see Fig.\ref{z0l}).  Vertical motions may replenish
  PAHs from below.
\label{zdisk.ps}}
\end{figure*}

\subsection{Disruption by Coulomb forces}

For completeness, we also mention disruption of PAHs by Coulomb
forces.  Double or multiple ionization of a PAH loosens the binding of
the peripheral H atoms as well as of the skeleton of carbon atoms.
The ejection of K-shell electrons by X-ray photons ($h\nu > 284$\,eV)
in combination with Auger electrons will amplify the process.  Coulomb
explosion is relevant mainly for small PAHs and neglected here.

\section{Conditions for PAH survival \label{surv}}

According to Eq.~(\ref{Stabil2}), PAHs are destroyed if the source
emits photons of energy $h\nu \ge 0.1\,N_c \,E_{\rm 0}$, irrespective of
the distance to the star or its luminosity.  For $N_c =100$, the
critical photon energy is only 50\,eV (Eq.\ref{Stabil2}).  As T Tauri
stars (or their jets) also radiate at X-rays and in the EUV, the
surface of the disk should be devoid of PAHs unless {\it a)} the
period over which hard photons are emitted is too short to destroy all
PAHs; {\it b)} the PAHs are by vertical motions removed from the hard
radiation before they are destroyed and there is an influx of PAHs
from below; {\it c)} PAH destruction is compensated by PAH formation
in the surface layer.  The last effect should, in a hard photon
environment where PAHs and carbon atoms are ionized, be prohibited by
Coulomb repulsion (Voit, 1992).

\subsection{Destruction time}

The above PAH survival condition under {\it a)} can easily be
dismissed.  To estimate the time for PAH removal, $t_{\rm rem}$, by
the radiation component $i$, we note that most of the radiation is
absorbed on the disk surface in a sheet of vertical optical depth
$\tau_\perp$ equal to the grazing angle $\alpha_{\rm g}$ of the
incident light.  We call this sheet the {\it{extinction layer}} (of
radiation component $i$) and denote its geometrical thickness $\ell_i$
(see Fig.\ref {zdisk.ps}).  To first order, the PAHs in the extinction
layer receive the stellar flux of Eq.(\ref {flux.eq}) with
$\tau_\nu=1$.  If the instability criterion of Eq.(\ref {Stabil2}) is
fulfilled, $t_{\rm rem}$ follows from $x N_\gamma t_{\rm rem} = N_c$,
where $x$ is from Eq.(\ref {x_Atome}), therefore
\begin{equation}\label{t_rem}
t_{\rm rem} \ = \ {N_c \over x} \ t_{\rm abs} 
\end{equation}

With $t_{\rm abs}$ from Table~\ref{dis.tab}, one sees that even at 100\,AU,
$t_{\rm rem}$ is short compared to the duration of the T Tauri phase
($\sim$$10^6$ yr, Bertout et al.~2007, Cieza et al. 2007).

\begin{figure*}
\begin{center}
\includegraphics[angle=0,width=12cm]{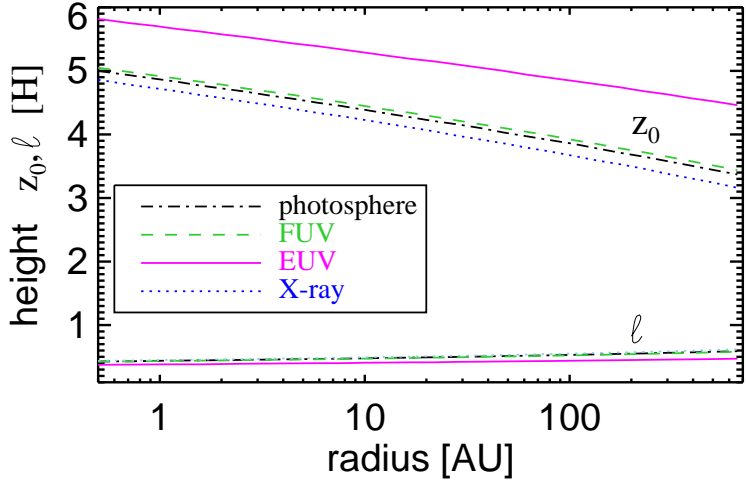}
\end{center}
\caption{\label{z0l} The height, $z_0$, of the bottom of the extinction layer
  and its thickness $\ell$ for the four radiation components (see
  Fig.\ref{zdisk.ps} and Eq.(\ref{z0.Eq}), (\ref{z0plusl})) for a grazing angle
  $\alpha_{\rm g} =3^{\rm o}$.  Due to the high gas densities, the EUV
  extinction layer is practically coincident with the ionization front and lies
  above the extinction layer of the other components.  This implies that EUV
  radiation is absorbed first whereas the other components penetrate deeper.
  Neglecting ionization by the FUV component, the gas below the $z_0$-line of
  the EUV radiation (top) is neutral. }
\end{figure*}

\subsection{Exposure time and vertical mixing}

Next we consider the possibility that vertical motions in the disk
lead to a continuous exchange between matter in the extinction layers,
where almost all photons are absorbed and PAHs destroyed, and the
layers below where PAHs are shielded and damaged ones possibly rebuilt
(Fig.\ref{zdisk.ps}).  We assume that gas and dust are perfectly mixed
in a mass ratio 130:1.

In a Keplerian disk that is isothermal in $z$-direction and in
hydrostatic equilibrium, the gas density changes like
\begin{equation}\label{rho_von_z}
\rho(z) = \sqrt{{2 \over \pi}} \ {\Sigma \over H} \ e^{-z^2/2H^2} 
\end{equation}
Here $\Sigma(r)$ is the surface density at radius $r$ which is assumed to follow
a power law,
\begin{equation}\label{Sigma_r}
\Sigma(r) = \int\limits_{0}^\infty \rho(z)\,dz = \Sigma_0 
\bigg[{r\over {\rm AU}} \bigg]^{-\gamma} 
\end{equation}
and

\begin{equation}\label{Skalen_H}
  H(r) = \sqrt{kTr^3/ GM_*m} 
\end{equation}
is the scale height, $M_* \simeq 1$\,M$_\odot$ the stellar mass and
$m$ the mass of a gas molecule.  For the surface density, reasonable
numbers are $\gamma =1$ and $\Sigma= 200$\,g\,cm$^{-2}$ (Hartmann et
al.~1998, Kitamura et al. 2002, Dullemond et al.~2002, Rafikov \&
Colle 2006, Gorti \& Hollenbach 2008), although the various estimates
show considerable scatter.

For the radial variation of the gas temperature in the opaque mid
plane, $T(r)$, we also adopt a power law,
\begin{equation}\label{T_von_r}
T(r) = T_0 \bigg[{r\over {\rm AU}} \bigg]^{-\beta} 
\end{equation}

The mid plane is roughly isothermal in $z$ because the optical depth
is high and the net flux zero.  It is much colder than the extinction
layers because it is not exposed to direct stellar heating.  The
radiative transfer in the disk, including the energy equation, can be
solved to any desired accuracy even when the disk is very opaque (see
section 11.3.2 of Kr\"ugel 2006).  As long as the dust in the mid
plane is optically thick to its own emission, the results for $T(r)$
can be well approximated by putting in Eq.(\ref{T_von_r}) $\beta =0.5$
and $T_0 \sim 130$\,K (as also suggested by Dullemond et al.~2007b or
Chiang \& Goldreich 1997).

Each extinction layer extends vertically from some value $z_0$ upwards
to infinity (Fig.\ref {zdisk.ps}).  We give it a finite thickness
$\ell$ by demanding that, say, 90\% of the photons are absorbed
between $z_0$ and $z_0+\ell$.  If $v_\perp$ denotes the typical
vertical velocity, for example, as a result of turbulence, PAHs are
exposed to radiation for a time
\begin{equation}\label{t_exp}
t_{\rm exp} = {\ell \over v_\perp} 
\end{equation}

This is also the mean residence time of a PAH in the extinction layer.
For PAHs to survive, $t_{\rm exp}$ must be smaller than $t_{\rm rem}$
Eq.(\ref{t_rem}).  The height $z_0$ follows from
\begin{equation}\label{z0.Eq}
K \int_{z_0}^\infty \rho(z)\, dz = \alpha_{\rm g}
\end{equation}
and $\ell$ may be estimated from the condition that only 10\% of the flux is
absorbed above $z_0+\ell$,
\begin{equation}\label{z0plusl}
K \int_{z_0+\ell}^\infty \rho(z) \, dz = 0.1 \ \alpha_{\rm g} 
\end{equation}

$K$ is the mass absorption coefficient of gas and dust at the
characteristic frequency of the particular radiation component (see
Table~\ref{dis.tab}).  Because $\rho(z)$ changes rapidly, $z_0$ is
rather insensitive both to $\alpha_{\rm g} $ as well as $K$.  For
$\alpha_{\rm g}/K=10^{-8}\ldots 10^{-2}$, one obtains $z_0=2.6H\dots
5.7H$.  So in the V band, where absorption is only by dust ($K\simeq
200$ cm$^2$ g$^{-1}$) and for a grazing angle $\alpha_{\rm g}= 3^{\rm
o}$, one gets $z_0\simeq (4\ldots 5)\, H$ and $f_\ell= \ell/H \simeq
0.5$. The height $z_0$ where an extinction layer begins and its
thickness are shown in Fig.\ref{z0l}; $\ell$ is for all radiation
components very similar ($\ell_i \sim H/2, \; i=1,\ldots,4)$.  EUV
photons are absorbed highest up, their extinction layer lies about one
scale height above the others.  X-rays penetrate slightly deeper than
photospheric or FUV photons.

\begin{figure*}
\begin{center}
\includegraphics[angle=0,width=12cm]{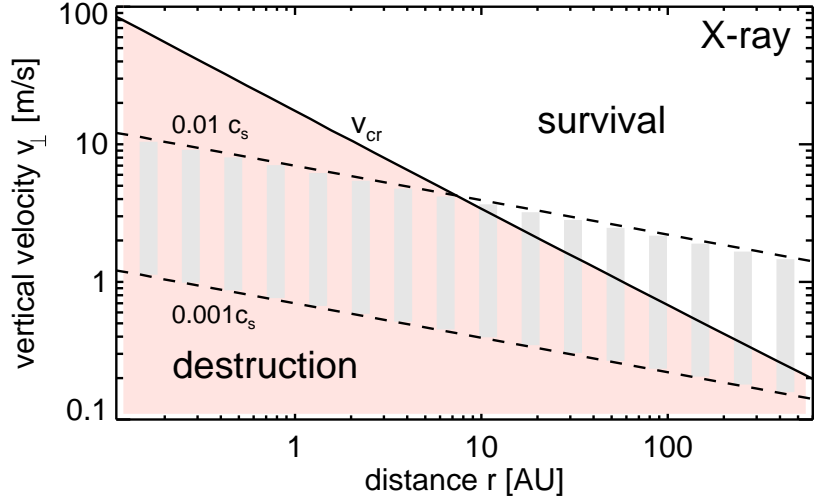}
\end{center}
\caption{The critical vertical velocity for PAH survival $v_{\rm cr}$
  after Eq.~(\ref{v_senk}) with respect to X--rays as a function of
  distance (full line).  We identify the vertical velocity $v_\perp$
  with the turbulent velocity $v_{\rm t}$. PAHs survive when $v_\perp
  = v_{\rm t} > \rm{v}_{\rm cr}$, else they are destroyed (shaded
  areas) by expulsion of atoms.  The hatched strip between the dashed
  lines shows the range where $v_\perp$ is between $0.001\, c_{\rm s}$
  and $0.01\, c_{\rm s}$ (where $c_{\rm s}$ is the sound velocity).
  \label{vperp.ps}}
\end{figure*}

\begin{table*}[htb]
 \caption{Quantities relevant to PAH survival. \label{dis.tab}  }
%\begin{center}
\begin{tabular}{|c| r c c c c | r r r r |}
  \hline
  &  (1) &   (2) & (3)  &    (4)   & (5) &  (6)  &(7) &(8) &(9) \\
\hline
  &   &    &   &      &  &    &  &  & \\
  component &$h\overline{\nu}$& $\kappa$ & $\eta$ & ${K/K_{\rm V}}$& $x$ 
  &$z_0/H$&$\ell/H$ 
  & $t_{\rm abs}$ & $v_{\rm cr}$ \\ 
  & eV & $10^{-4}$\AA$^2$ & & & & &&  & m/s \\
  \hline
  photosphere & 2.7& 700 &$\ll 1$ & 1.2 & -  & 4.4 &0.5& 5 s  & - \\
  FUV         & 4.2& 700&$\ll 1$ & 1.5 & -  & 4.5 &0.5&  120 s  & - \\
  EUV         & 97       & 36  &$\sim$0.5&110 & 9   &5.3&0.4&9\,days&2900 \\
  X--ray      & 1100  & 0.17 &$\ll 1$& 0.55 &100 & 4.2 &0.5& 290\,yr &3.4 \\
  \hline
\end{tabular}
%\end{center}

\vspace{0.5cm}

(1) The mean energy of destructive photons, $h\overline{\nu}$, from
Eq.(\ref{hnu_mean}) at the bottom of the extinction layer (see
Fig.\ref {zdisk.ps}); for photospheric and FUV photons we put
$\nu_{\rm c} =0$ and for EUV and X-rays we integrate for $\nu \geq
\nu_{\rm c}$ (Eq.~\ref{Stabil2});

 (2) absorption cross section, $\kappa$, per C atom at
frequency $\overline{\nu}$; 

 (3) approximate mean degree of ionization of the gas in
the extinction layer; 

 (4) extinction cross section of gas and dust at frequency
$\overline\nu$ normalized to $K_{\rm V}= 200$ cm$^2$ g$^{-1}$; 

 (5) number of expelled atoms, $x$, per absorption event
from Eq.(\ref{x_Atome}); 

 (6) altitude of the bottom of the extinction layer in
units of the scale height $H$; 

 (7) thickness of extinction layer; 

 (8) mean time $t_{\rm abs}$ from Eq.(\ref{t_abs}) in
which one photon is absorbed; 

 (9) critical vertical velocity for PAH survival.  

 Note: $z_0, \ell, t_{\rm abs}$ and $v_{\rm cr}$ refer
to $r=10$\,AU.

\end{table*}

\begin{figure*}
\begin{center}
\includegraphics[angle=0,width=12.cm]{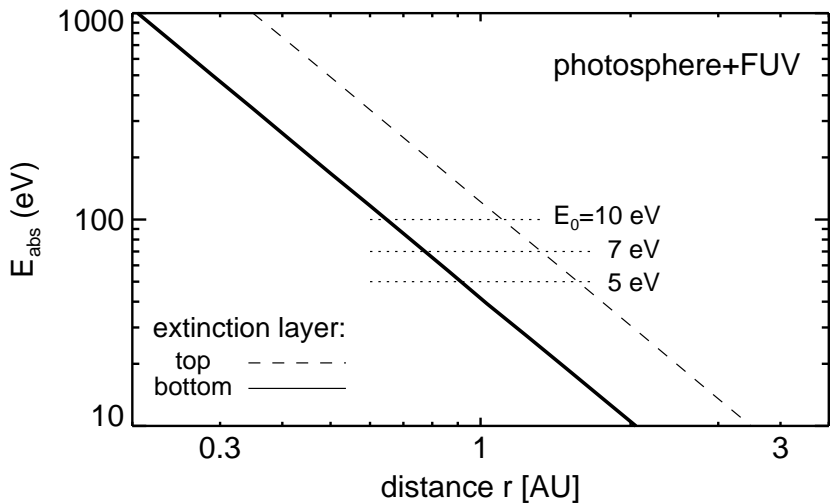}
\end{center}
\caption{The {energy $E_{\rm {abs}}$ (Eq.\ref{E_abs}), which is
    absorbed by a PAH of $N_c=100$ carbon atoms, as a function of
    distance from the star.  The PAH is exposed to the photospheric
    and FUV radiation component described in Table~\ref{spec.tab}. The
    dashed line refers to the top of the extinction layer ($\tau=0$)
    and the full line to its bottom ($\tau=1$).  For Arrhenius energy
    of $E_0= 5$, 7 and 10\,eV, the minimum energy input $\Delta E$
    (Eq.\ref{Stabil2}) for PAH dissociation is indicated by the dotted
    lines.}
  \label{Eabs.ps}}
\end{figure*}

From Eq.(\ref{t_abs}), (\ref{t_rem}) and (\ref{t_exp}), one finds that
vertical motions safeguard PAHs against destruction if
\begin{equation}\label{v_senk}
v_\perp > v_{\rm{cr}}= {\ell x \over t_{\rm abs} N_c} = {f_\ell
Hx\over 4\pi r^2} \int\limits_{\nu_{\rm c}}^\infty {L_\nu
e^{-\tau_\nu} \kappa_\nu\over h\nu}\,d\nu
\end{equation}

with $\nu_{\rm c}$ from Eq.(~\ref{Stabil2}).  When the scale height
$H$ is given by Eq.(\ref{Skalen_H}), $v_{\rm{cr}} \propto
r^{-(\beta+1)/2} = r^{-3/4}$.  The values of $v_{\rm{cr}}$, at a
distance of 10\,AU, are listed in Table~\ref{dis.tab} together with
other quantities relevant to PAH survival.

{One expects the disk also to be turbulent.  Turbulence may be
  driven by various processes such as shear flows in the disk (Lin \&
  Bodenheimer 1982), magneto-rotational instabilities (Balbus \&
  Harley 1991) or velocity discontinuities at places where infalling
  matter (Cassen \& Mossman 1981) or outflows (Elmegreen 1978) strike
  the disk surface.  Until now it is not clear which type of
  turbulence dominates. We assume that for the size of the largest
  Eddies, $\ell_{\rm{ed}}$, the average turbulent velocity, $v_{\rm
    t}$, grows linearly with the sound speed $c_{\rm s}$. Various
  hydrodynamical 3--dimensional calculations (Cabot 1996, Boss 2004,
  Johansen \& Klahr 2005, Fromang \& Papaloizou 2006) support this
  view. The favoured parametrisation is $v_{\rm t} = \alpha^{q} \/
  c_s$ with $q = 0.5$.  This choice has consequences on the Eddy scale
  $\ell_{\rm {ed}} = \alpha^{1-q} \/ H$ and the turn over time $t_{\rm
    {ed}} = \ell_{\rm {ed}} / v_{\rm t} = \alpha^{1-2q} / \Omega_{\rm
    K}$, with Keppler frequency $\Omega_{\rm K} = \sqrt{G M_{*}/ R^3}$
  (e.g. Dullemond \& Domink 2004).  Weidenschilling \& Cuzzi (1993)
  use $q = 1$ so that the Eddy scale is about the pressure scale
  height, $\ell_{\rm {ed}} = H$, and larger than the thickness of the
  extinction layer, $\ell_{\rm {ed}} > \ell \sim H/2$. Estimated
  values for $\alpha$ are in the range from 0.0001 up to 0.1
  (Dullemond \& Dominik 2004, Schr\"apler \& Henning 2004, Youdin \&
  Lithwick 2007). Taking $\alpha = 0.01$ the Eddy scale is 10 times
  smaller for $q = 0.5$ than for $q = 1$, and in addition larger
  turbulent velocities are obtained with $v_{\rm t} = \sqrt{\alpha} \/
  c_s$, supporting a faster transport of the PAH.}  Identifying
$v_\perp$ in Eq.(\ref{v_senk}) with the turbulent velocity $v_{\rm t}$
and assuming a temperature dependence as in Eq.(\ref {T_von_r}), we
plot in Fig.\ref{vperp.ps} the vertical velocity $v_\perp$ as a
function of radius.  The figure also shows the critical velocity for
PAH survival, $v_{\rm cr}$, with respect to X--rays.  {Note that
  the critical velocity is for the X--ray radiation component
  insensitive to the particular choice of $E_0$ and $\nu_c$
  (Eq.~\ref{Stabil2}, ~\ref{v_senk}).} When $v_\perp = 0.01 \, c_{\rm
  s}$, PAHs can survive at distances $r > 10$ AU; when $v_\perp$ is
considerably smaller than $0.01 \, c_{\rm s}$, they cannot.

Critical velocities for PAH survival are much higher for EUV than for
X--ray photons (Table~\ref{dis.tab}).  EUV radiation will therefore
always destroy PAHs but, as depicted in Fig.\ref{z0l}, the EUV
extinction layer is the topmost and below it, PAHs may survive and be
excited.

If PAHs are removed from the extinction layer before they are
destroyed, they must, in order to be detected, at the same rate be
injected from below.  Therefore, the critical velocity can
alternatively be expressed through
\begin{equation} \label{vin}
{1\over t_{\rm rem}} \int_{z_{\rm {0}}}^{z_0 + \ell } \rho(z) \, dz  
\simeq \rho(z_0) \, v_{\rm cr}
\end{equation}
which leads to similar values.  We note that the mass reservoir below the
extinction layer is sufficient to sustain, over the lifetime of the disk $t_{\rm
  {life}}$, the required mass influx $\rho v_{\rm cr}$.

\subsection{PAH dissociation by soft versus hard photons \label{comp.sec}}

\noindent 
{The energy $E_{\rm {abs}}$ (Eq.\ref{E_abs}), which is absorbed by
  a PAH of $N_c=100$ carbon atoms, is shown in Fig.~\ref{Eabs.ps} as a
  function of distance from the star.  The PAH is exposed to the
  photospheric and FUV radiation component described in
  Table~\ref{spec.tab} and results are shown for the top ($\tau=0$)
  and the bottom ($\tau=1$) of the extinction layer.  The minimum
  energy input $\Delta E$ (Eq.\ref{Stabil2}) for PAH dissociation
  depends on the choice of the Arrhenius energy $E_0$ and is indicated
  for $E_0= 5$, 7 and 10\,eV, respectively.  In this picture, for
  $E_0= 5$\,eV and at the bottom of the extinction layer, PAHs are
  dissociated by soft photons up to 1\,AU. For X--rays, however, we
  find that PAH destruction occurs typically at distances up to $\sim
  10$\,AU or even larger (Fig.\ref {vperp.ps}).  Dissociation of PAH
  acts for soft (photospheric and FUV) photons on much shorter
  distances than for hard photons (X--ray component). }

\section {Large grains}

Observations of T Tauri stars at millimeter wavelengths (Testi et
al. 2003, Lommen et al. 2007) and in the mid infrared (van Boekel et
al. 2003, Przygodda et al., 2003, Kessler-Silacci et al. 2006, Bouwman
et al.~2008, Watson et al. 2009) suggest that grains in T Tauri disks
are at least 10 times larger than those in the ISM. As such large
grains may also be present in the top disk layer we estimate how this
would affect the stability analysis of PAHs.  We first note that in
case of homogeneous mixing an increase in particle size would not
alter the dust--to--gas mass ratio.

Should the grains be much larger than the wavelength, the absorption
coefficient per gram of dust, $K_{\rm d,\lambda}$, would decrease
roughly like one over grain radius whereas the ratio $K_{\rm
d,\lambda}/ K_{\rm d,V}$ would still roughly be given by the values in
Table~\ref{dis.tab}.  For hard X--rays, on the other hand, $K_{\rm
d,\lambda}$ is not sensitive to grain size.

Therefore, if disk grains are on average ten times bigger and thus
$10^3$ more massive than interstellar ones, we expect that the height
$z_0$ to which the stellar radiation components can penetrate (see
Fig.\ref {z0l}) stays the same for X--rays, but also for EUV radiation
because EUV absorption is due to gas, not dust.  However, optical and
FUV photons will reach farther down, about half a scale height, so
that there may be a thin disk layer ($\sim$$H/4$) where PAHs are
shielded from X--rays and EUV photo-destruction and excited by optical
or FUV radiation.

\section {Conclusion}\label{conclusion}

In the search for an explanation why most T Tauri stars do not exhibit PAH
features and only a few do, we investigate which processes can remove PAHs from
the surface layer of T~Tauri disks and under which conditions they should be
present.  Clearing of PAH through interaction with planets seems not an
efficient process (Geers et al 2007b) and we show that PAH under--abundance can
be caused by radiative destruction. We use a fiducial model for the photon
emission of the T Tauri star that includes, beside the photosphere, FUV and EUV
radiation and an X--ray component.

\begin{enumerate}
\item We introduce for each stellar radiation component the notion of {\it
    extinction layer} as the place where $\sim$90\% of the photons are absorbed.
  EUV photons are mainly absorbed by gas, X-rays by gas and dust alike and the
  photospheric and FUV component are only attenuated by dust.  The extinction
  layer of all four components have a similar geometrical thickness and their
  bottom is at similar elevation $z_0$, except for the EUV extinction layer
  which lies higher up (Fig.\ref{z0l}).

\item PAH may be radiatively destroyed, {by unimolecular
  dissociation, where one or several atoms are expelled after photon
  absorption.

\item Destruction by the photospheric and FUV radiation component
  (soft photons), increases with the strength of the radiation field
  and is very efficient below 1\,AU.

\item Hard photons can dissociate PAHs} at all distances and their
  efficiency grows with the hardness of the photons.  Without some
  counter process, all PAHs (in layers where they can be excited)
  would be destroyed within a time short compared to the lifetime of
  the disk.

\item Although grains in the disk surface are presumably larger than
  interstellar ones, the stability analysis of PAHs would not change
  significantly.

\item Therefore, in disks where PAHs are detected, there must be some survival
  channel.  Because creation of PAHs in the extinction layer is too slow to
  compete with PAH destruction (Voit 1992), we suggest {\it vertical mixing} as
  a result of turbulence.  It can replenish PAHs or remove them from the reach
  of hard photons.

\item For standard disk models, the minimum velocity for PAH survival
  is proportional to $r^{-3/4}$ and equals $\sim$5\,m/s at 10\,AU.  If
  turbulent velocities are proportional to the sound speed a velocity
  $v_{\rm t} \ge 5$ m/s would imply $v_{\rm t} / c_{\rm s} \,
  \simgreat \, 0.01$ as PAH survival condition.  Theoretical
  predictions for this ratio have a large spread but in accordance
  with the observational fact that PAH features are usually absent it
  seems that generally the condition is not fulfilled.

\item A higher PAH detection rate is found in Herbig Ae/Be stars.{
  In our picture this is explained} as their destructive hard
  radiation component is relatively weak ($L_{x}/L_{*} \simless
  10^{-7}$, Preibisch et al.\,2006) and because the intensity of the
  PAH emission from large distance from the star is larger given their
  higher {optical} luminosities.

\end{enumerate}

\begin{acknowledgements} 
{We thank the second anonymous referee for constructive comments.}
\end{acknowledgements}

\end{document}